\documentclass[lettersize,journal]{IEEEtran}
\usepackage{amsmath,amsfonts}
\usepackage{algorithmic}
\usepackage{algorithm}
\usepackage{array}
\usepackage[caption=false,font=normalsize,labelfont=sf,textfont=sf]{subfig}
\usepackage{textcomp}
\usepackage{stfloats}
\usepackage{url}
\usepackage{verbatim}
\usepackage{graphicx}
\usepackage{cite}
\usepackage{pgfplots} 
\usepackage{pgffor}
\pgfplotsset{compat=1.16}
\usepgfplotslibrary{colorbrewer}
\pgfplotsset{cycle list/Set1-4}

\usepgfplotslibrary{fillbetween}
\begin{document}

\title{Spectro-temporal unitary transformations for coherent modulation: design trade-offs and practical considerations}

\author{Callum~Deakin, Xi Chen
\thanks{The authors are with Nokia Bell Labs, Murray Hill, NJ 07974, USA. (e-mail: callum.deakin@nokia-bell-labs.com).}
\thanks{Manuscript compiled \today.}}

\markboth{\MakeUppercase{\today}}%
{Shell \MakeLowercase{\textit{et al.}}: A Sample Article Using IEEEtran.cls for IEEE Journals}


\maketitle

\begin{abstract}
This paper analyzes the performance of spectro-temporal unitary transforms for coherent optical modulation. Unlike conventional IQ modulation, such transforms are based on a cascade of phase modulators and dispersive elements, so are theoretically lossless and not limited by the bandwidth of the constituent modulators. We analyse the performance limits and design trade-offs of this scheme: estimating how the number of stages, amount of dispersion, modulator bandwidth, symbol block length and electrical signal power impacts the achievable signal-to-distortion ratio (SDR). Importantly, we show that high ($>30$~dB) SDRs suitable for modern $>200$~GBd class coherent optical communications are achievable with a low ($<6$) number of stages and reasonable parameters for driver power, modulator bandwidth and on-chip dispersion. Finally we address the SDR penalties associated with potential phase, amplitude, or dispersion errors, and limited DAC resolution.
\end{abstract}


\section{Introduction}
Conventional coherent modulation is based on amplitude modulation via two $\pi/2$ out-of-phase Mach-Zehnder modulators (MZM) to achieve continuous arbitrary manipulation of the optical carrier, as shown in Fig.~\ref{experiment_setup}(a). This operation is fundamentally lossy and bandwidth limited, since it is based on amplitude modulation via switching: the excess light in low amplitude symbols is dumped in the terminated output port of the MZM and can only be switched at a speed that is bounded by the bandwidth of the constituent MZMs. 

Constant improvement in transmitter baud rates has historically been essential to reduce the cost per bit and power efficiency of optical transceivers, with current demonstrations exceeding 400~GBd~\cite{che2026440}. With this in mind, it is conceivable that the future networks may require single transceivers with bandwidths in excess of 1 THz that will operate in a network with both wavelength and spatial parallelism~\cite{winzer2017scaling,schmogrow2022solving}.  It is unclear whether electronic integrated circuit technology (namely the DACs, drivers, and modulators) can support this future when the $f_{max}$ of even the highest frequency semiconductors such as InP HBT is barely exceeding 1~THz~\cite{urteaga2017inp}.

At the same time, the optical loss of conventional IQ modulators is exacerbated by the nonlinear MZM response and high peak to average power ratio of typical $M$-ary QAM coherent signals. In particular, aggressive pulse shaping and probabilistic constellation shaping to achieve maximum spectral efficiency often results in modulation losses in excess of 25~dB. This problem may become critical in future transceivers as bandwidths approach 1~THz since the required optical power to overcome shot noise at these bandwidths is unachievable by a single integrated laser source and would dominate the transceiver power budget. Furthermore, the deployment of hollow-core fibers in future networks may allow linear operation at high transmitter output powers to maximise fiber capacity and reach, further necessitating high transceiver output power~\cite{liu2018nonlinearity,poggiolini2022opportunities}. 

An alternative modulation scheme based on multiple cascaded stages of phase modulators and dispersive elements has been proposed and demonstrated~\cite{thiel2017programmable,mazur2019optical}, as shown in  Fig.~\ref{experiment_setup}(b). The scheme achieves amplitude modulation by redistributing light temporally rather than switching, and is both theoretically lossless and unconstrained by the bandwidth of the constituent phase modulators. It has been further shown that this scheme can be operated in real time~\cite{saxena2023performance} and can modulate different wavelengths independently without a mux/demux. The lossless nature of the scheme also makes it particularly attractive for temporal mode sorting in quantum information processing~\cite{ashby2020temporal}.

This paper explores the hardware requirements to implement this scheme in a coherent transmitter through numerical simulations as an extension of our conference paper~\cite{deakinofc2025}. In section 2 we describe in detail the theoretical basis for arbitrary waveform generation using spectro-temporal unitary transformations under given hardware constraints, describe the multi-objective optimisation procedure used in this paper and derive the required gradient functions. Section 3 presents results that estimate the theoretical limits of these transformations and explores design trade-offs, while section 4 examines how practical limitations and implementation errors degrade the achievable SDR.

\section{Theory and operating principle}

\begin{figure*}[tb]
   \centering
    \includegraphics[width=0.8\linewidth]{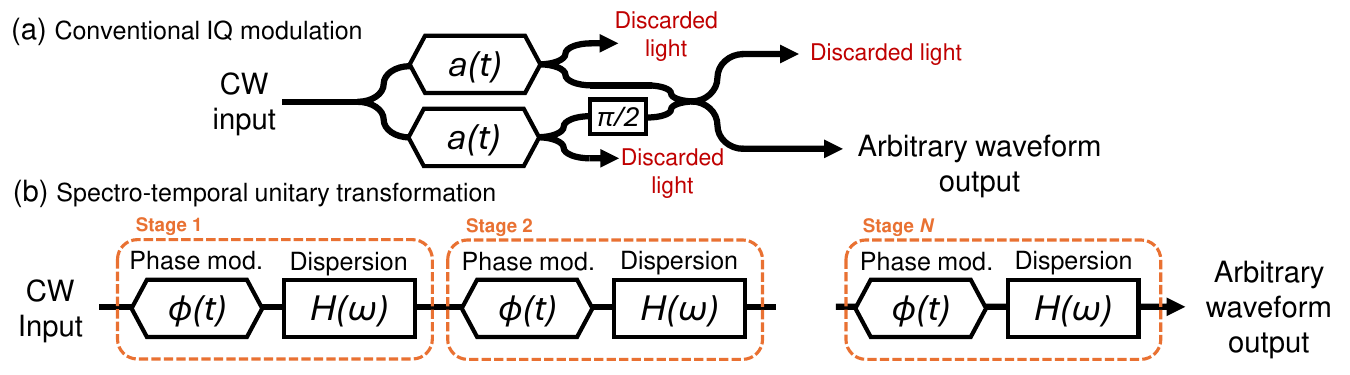}
    \caption{(a) Conventional IQ modulation based on amplitude modulation, $a(t)$. (b) Lossless spectro-temporal unitary transform based arbitrary waveform modulation based on cascaded phase modulators $\phi(t)$ and dispersive elements $H(\omega)$.}
    \label{experiment_setup}
\end{figure*}

A linear optical device can be represented by the operator $U$, which transforms a vector describing the input modes $x_m$ into a vectors of output modes $y_m$ in a particular basis set

\begin{align} \label{U_framework}
     \underbrace{
     \begin{pmatrix}
           y_{1} \\
           y_{2} \\
           \vdots \\
           y_{m}
     \end{pmatrix}}_{\text{Output}} 
     &= 
     \underbrace{
     \begin{pmatrix}
            u_{11} & u_{12} & \cdots & u_{1m}\\
            u_{21} & u_{22} & \cdots & u_{2m}\\
            \vdots & \vdots & \ddots & \vdots\\
            u_{m1} & u_{m2} & \cdots & u_{mm}
     \end{pmatrix}}_{U}
     \underbrace{
     \begin{pmatrix}
           x_{1} \\
           x_{2} \\
           \vdots \\
           x_{m}
     \end{pmatrix}}_{\text{Input}}
\end{align}

where $U$ is a unitary matrix. To achieve arbitrary transformation between the input and output modes, we must be able to construct every unitary matrix $U$. In the context of optics, `mode' could mean spatial, spectral or temporal modes. For example, pixels or Laguerre–Gaussian modes as spatial modes, or discrete time samples or Hermite–Gaussian modes as temporal modes. In this paper we will focus on discrete time samples as temporal modes, but the framework described here can be applied to any orthonormal basis that is able to describe optical wave propagation in time or space.

The key question that arises in the framework described by~(\ref{U_framework}) is: how can we construct every possible transformation matrix $U$? The answer lies in decomposing the desired matrix $U$ into physically realizable sub-matrices. Proposition 7 in~\cite{schmid2000decomposing} describes how, as a consequence of~\cite[Lemmas 2 and 3]{borevich1981subgroups}, a sub-group of matrices containing all unitary diagonal matrices and a matrix $H$ which has all its entries non-zero is in fact the full group of unitary matrices. In other words, all unitary matrices can be built with a succession of variable diagonal matrices $\Lambda_n$ and a fixed matrix $H$ with no non-zero entries 

\begin{equation}
    U = \Lambda_1H\Lambda_2H\Lambda_3H \cdots \Lambda_nH
\end{equation}

where
\begin{equation}
\Lambda_n = \begin{pmatrix}
            e^{i\phi_1} & 0 & \cdots & 0\\
            0 & e^{i\phi_2} & \cdots & 0\\
            \vdots & \vdots & \ddots & \vdots\\
            0 & 0 & \cdots & e^{i\phi_m}
     \end{pmatrix}
     \end{equation}
\begin{equation}
H = \begin{pmatrix}
            h_{11} & h_{12} & \cdots & h_{1m}\\
            h_{21} & h_{22} & \cdots & h_{2m}\\
            \vdots & \vdots & \ddots & \vdots\\
            h_{m1} & h_{m2} & \cdots & h_{mm}
     \end{pmatrix}
     \end{equation}

The matrix $H$ represents a mode-mixing operation in which every output mode is a linear composition involving every input mode. An example of a matrix which fulfils this condition is the discrete Fourier transform matrix
\begin{equation}
    \textnormal{DFT}_m = \frac{1}{\sqrt{m}}\Big(\exp{\frac{2\pi i}{m} \Big)}^{j\cdot k}_{j,k=0,\dots, m-1}
\end{equation}

for an $m\times m$ matrix. Therefore, it is proven~\cite{schmid2000decomposing,morizur2010programmable} that every unitary matrix can be represented as a product of diagonal matrices and the discrete Fourier transform matrix. In fact, although any matrix with no non-zero entries can be used as $H$, it has been shown that the complex Hadamard matrices, of which the DFT is an example, provide optimal mode-mixing~\cite{alvarez2025universality}.

The optical transformation represented by $U$ may describe a transformation on spatial or temporal modes. It is well known that free space propagation and linear chromatic dispersion can approximate a Fourier transform on spatial and temporal modes respectively~\cite{kolner1994space}. For linear chromatic dispersion in the time domain, this condition arises when
\begin{equation} \label{dispersion_condition}
    |\beta_2 L|\gg \tau_w^2
\end{equation} 
for group delay dispersion $\beta_2 L$ and where $\tau_w$ is the temporal width of the pulse~\cite{salem2013application}. This is a general condition resulting from the diffusion heat equation~\cite{kolner1994space}, which simply states that if every mode experiences a sufficiently different linear group delay, then the individual modes will become separated. On the other hand, a unitary diagonal matrix $\Lambda_n$ simply represents a independent phase modulation of the individual modes (i.e. time samples). Therefore, implementing this transformation in an optical system can be achieved by a cascade of phase modulators and either free space propagation (spatial modes) or chromatic dispersion (temporal modes). The temporal formulation, which is the focus of this paper, is shown in Fig.~\ref{experiment_setup}(b).

This method was first described and implemented on spatial modes~\cite{morizur2010programmable}, the realization of which is known as a multi-plane light wave converter (MPLC). Theoretical work has shown that a necessary condition to be able create any unitary (i.e. `universality') is that $N\geq M+1$ for $N$ stages and matrix dimension $M$~\cite{alvarez2025universality}, which is supported by numerical evidence~\cite{saygin2020robust,zelaya2024goldilocks} and the existence of deterministic algorithms~\cite{lopez2021arbitrary,girouard2025near}. It has also been suggested that even the fractional Fourier transform~\cite{ozaktas2001fractional} can provide sufficient mixing for practical implementation~\cite{zelaya2024goldilocks,alvarez2025universality,markowitz2023universal}, which describes the case of sub-Fraunhofer distance (i.e. Fresnel, or `near field') diffraction~\cite{ozaktas2001fractional} or when the condition~(\ref{dispersion_condition}) is not fulfilled.

In this paper however, we do not require complete universality and are interested only in the subset of transforms that describes the transformation of a continuous wave laser into a typical coherent modulation signal with sufficient fidelity: i.e. RRC shaped $M$-ary QAM signals encoding blocks of a continuous random bit stream. In this case, the temporal modes are simply the time samples in a discrete time system. Restricting the required set of unitaries to this subset (or any other) results in substantially fewer required stages $N$, as has been shown by experimental results~\cite{mazur2019multi,fontaine2019laguerre}.

\subsection{Optimization}

Although we have shown that any transformation (and therefore any arbitrary waveform) can be produced from a cascade of phase modulators and dispersion, finding a set of phase instructions (or modulations) $\{\phi_1(t),\phi_2(t),\dots \phi_N(t)\}$ that produce a specific waveform is non-trivial. Analytical solutions for the phase instructions required to achieve an arbitrary transformation are known only for low mode/phase mask counts~\cite{yasir2025compactifying} (or the trivial case of $N=1$~\cite{wyrowski1991upper,verhoeven2016upper}), therefore it is common to use numerical optimisation to find the required phase manipulations to produce a particular transform.
\begin{figure*}[tb]
   \centering
    \includegraphics[width=0.8\linewidth]{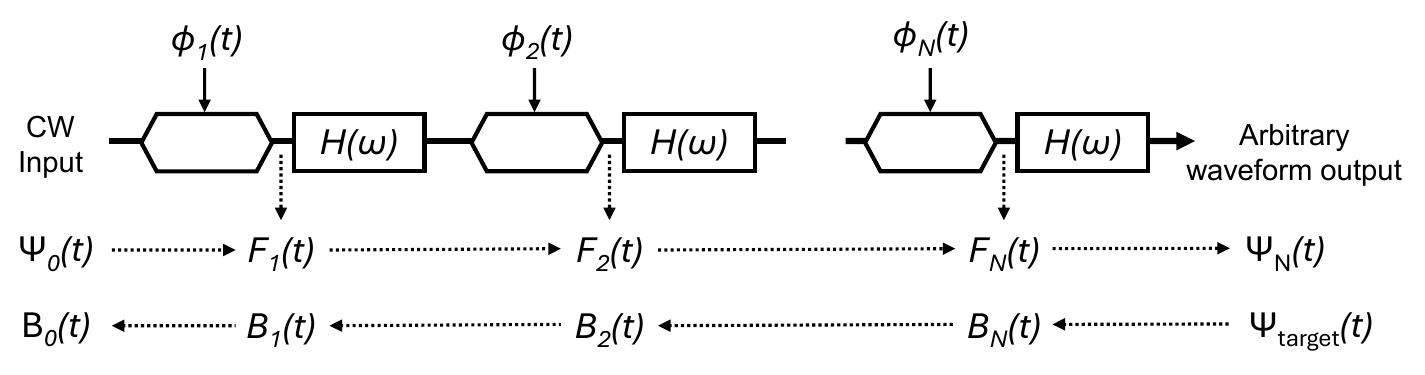}
    \caption{The optimisation procedure attempts to find the set of phase instructions, $\{\phi_1(t),\phi_2(t),\dots \phi_N(t)\}$, at each stage that can produce a waveform $\Psi_N(t)$ after $N$ stages that matches the target waveform $\Psi_\textnormal{target}$ with minimal DSR. The DSR gradient is calculated using the forward propagated wave $F_n(t)$ and backward propagated wave $B_n(t)$ at each stage. The backward propagating wave is calculated by propagating the target waveform $\Psi_\textnormal{target}$ backwards through the system, while the forward propagating wave is calculated by propagating the input waveform $\Psi_0$ forwards through the system, using the current phase instructions.}
    \label{optimsation_setup}
\end{figure*}
Our objective is to minimise the distortion-to-signal ratio (DSR), or equivalently maximise the signal-to-distortion ratio (SDR), as referenced to the target waveform, $\Psi_\textnormal{target}(t)$. We follow the derivation in~\cite{saxena2023performance} with slight modifications. The distortion to signal ratio (DSR) is given by 

\begin{equation}\label{DSR_definition}
    \textnormal{DSR} = \frac{1}{E_\textnormal{target}}\sum_{m=0}^M |\Psi_\textnormal{target}(mT) - \Psi_N(mT)|^2
\end{equation}
for $N$ stages operating on a input vector of length $M$, where
\begin{equation}
   E_\textnormal{target} = \sum_{m=0}^M |\Psi_\textnormal{target}(mT)|^2
\end{equation}
and $\Psi_N(mT)$ is the generated waveform after $N$ stages. All vectors represent a time sampled system with sampling period $T_s$. Expanding (\ref{DSR_definition}) gives
\begin{multline}
    \textnormal{DSR} = \frac{1}{E_\textnormal{target}}\sum_{m=0}^M (\Psi_\textnormal{target}^*(mT) - \Psi_N^*(mT)) \times \\ (\Psi_\textnormal{target}(mT) - \Psi_N(mT))
\end{multline}
Since all operations are unitary, we can express the DSR as the difference between the forward and backward propagating waves, $F_n(t)$ and $B_n(t)$ respectively, as shown in Fig.~\ref{optimsation_setup}. This formulation highlights the unitary nature of the transformation and allows us to derive the derivative of the DSR with respect to the change in phase modulations at each stage. In other words, it shows us exactly how each stage changes the waveform. After the $n$-th stage
\begin{equation}
\textnormal{DSR} = \frac{1}{E_\textnormal{target}}\sum_{m=0}^M |B_n(mT) - F_n(mT)|^2.
\end{equation}
The forward waves $F_n(t)$ are generated from the input CW waveform, $\Psi_0(mT)$ while the backward waves $B_n(t)$ are generated from the target waveform $\Psi_\textnormal{target}(t)$, as illustrated in Fig.~\ref{optimsation_setup}. This expands to
\begin{multline} \label{DSR_unconstrained}
\textnormal{DSR} = \frac{1}{E_\textnormal{target}}\sum_{m=0}^M (B_n^*(mT) - F_n^*(mT))\times \\ (B_n(mT) - F_n(mT)).
\end{multline}
To find the gradient, we differentiate the DSR with respect to the phase instructions $\phi_n(mT)$. Noting that 
\begin{equation}
    \frac{\partial  F_n(mT)}{\partial \phi_n} = iF_n(mT)
\end{equation}
\begin{equation}
    \frac{\partial  B_n(mT)}{\partial \phi_n} = 0
\end{equation}
we find that
\begin{multline} 
    \frac{\partial \textnormal{DSR}}{\partial \phi_n} =  \frac{1}{E_\textnormal{target}}\sum_{m=0}^M i(F_n^*(mT)B_n(mT)\\- F_n(mT)B_n^*(mT))
\end{multline}
\begin{equation} 
    \frac{\partial \textnormal{DSR}}{\partial \phi_n} =  \frac{1}{E_\textnormal{target}}\sum_{m=0}^M -2\textnormal{Im} (F_n^*(mT)B_n(mT))
\end{equation}
\begin{equation} \label{gradient_unconstrained}
    \frac{\partial \textnormal{DSR}}{\partial \phi_n} =  \frac{2}{E_\textnormal{target}}\sum_{m=0}^M \textnormal{Re} (iF_n^*(mT)B_n(mT))
\end{equation}
Since we have obtained an analytical expression for the gradient, we can use any number of gradient-descent optimisation approaches. Our previous work~\cite{deakinofc2025} and other papers~\cite{mazur2019multi} instead used a modified wavefront matching algorithm that propagated the input CW light and target waveform back and forth through the system while minimising the phase difference between the backwards and forwards propagating wave at each stage~\cite{sakamaki2007new}. Although effective at finding solutions, this method introduces an implicit power constraint since as it attempts to minimise the phase change at each stage, and does not allow the imposition of additional optimisation objectives as will be discussed in the next section. Other possible approaches include simulated annealing~\cite{ashby2020temporal}. 

The gradient-descent based approach used in this paper is therefore more flexible and better able to explore the global set of solutions subject to any constraints we may impose. Specifically, we use the limited memory Broyden–Fletcher–Goldfarb–Shanno algorithm with simple box contraints (L-BFGS-B)~\cite{byrd1995limited}. This is a second-order gradient descent method that makes an implicit approximation of the Hessian matrix using a limited number of past updates, and it particularly well suited to problems with a large number of variables. The memory size (i.e. the number of past updates stored) in this case is set as 10. The convergence criterion is set as

\begin{equation}
    \frac{|\textnormal{DSR}_{k+1} - \textnormal{DSR}_k|}{\max(1, |\textnormal{DSR}_k|)} \leq \epsilon_{64}
\end{equation}

where $\epsilon_{64}$ is the machine precision of the 64-bit floating point numbers used in these simulations, i.e. $2^{-52}$. 

An unavoidable constraint in optical transceivers is that the available output power of the modulator drivers is limited. Therefore to find  practical set of phase instructions we may wish to simultaneously minimise the average power of the drive instructions 

\begin{equation}
    P_n = \frac{1}{M}\sum_{m=0}^M \phi_n^2({mT})
\end{equation}
which has derivative
\begin{equation}
     \frac{\partial P_n}{\partial \phi_n} = \frac{1}{M}\sum_{m=0}^M 2\phi_n({mT})
\end{equation}

We can combine these 2 objectives via linear scalarization to form a single objective function
\begin{equation} \label{multi_objective_1}
    f(mT) = \textnormal{DSR}(mT) + aP_n(mT)
\end{equation}
for scalarization parameter $a$
\begin{multline} \label{multi_objective}
    f(mT) = \frac{1}{E_\textnormal{target}}\sum_{m=0}^M |\Psi_\textnormal{target}(mT) - \Psi_N(mT)|^2 \\+ \frac{a}{M}\sum_{m=0}^M \phi_n^2({mT})
\end{multline}
since differentiation is a linear map, we can simply sum the gradients

\begin{multline} 
    \frac{\partial f(mT)}{\partial \phi_n} =  \frac{2}{E_\textnormal{target}}\sum_{m=0}^M \textnormal{Re} (iF^*(mT)B(mT)) \\+ \frac{2a}{M}\sum_{m=0}^M \phi_n({mT})
\end{multline}
The scalarization parameter $a$ is constant that determines the relative weight of the objectives in the optimisation procedure. $a$ can therefore be adjusted until the desired balance between modulator drive power and SDR is achieved. If the optimisation results in a set of phase instructions that have too high an average power, then $a$ can be increased to reduce the average power at the cost of reduced SDR. Since the outcome set is convex, i.e. increasing SDR or decreasing drive power always reduces optimality in the other objective, this will yield a Pareto optimal solution for the drive power and SDR.

\section{Performance and design trade-offs}

\subsection{Unconstrained RF drive power}

\begin{figure}[t]
   \centering
    \input{figs/fig1}
    \caption{SDR v. dispersion per stage ($\beta_2 L$) for a $B_{\textnormal{PM}}=$ 0.55~$f_s$, and varying number of stages, $N$. The dispersion is normalised to the symbol period squared ($T_s^2$) and shaded areas indicate minimum and maximum values.  Example constellations and power spectral densities of the converged phase modulations inset in (a) and (b).}
    \label{fig:dispersion}

\end{figure}

To understand the performance of the modulation technique for coherent optical communications without loss of generality, we perform the describe optimization for root raised cosine shaped (shaping factor $\beta=0.1$, 10\% roll-off) 16-QAM and compare the result to the desired target waveform by calculating the signal-to-distortion ratio (SDR). First we perform phase instruction optimisation without any power constraints in order to estimate the ultimate limits of the modulation technique. The simulation is oversampled by a factor of eight (i.e. $\frac{1}{T}$ = 8~$f_s$) since simulating at a lower sampling rate can result in erroneously high SDR estimations: aliasing as higher frequencies are generated (either by strong phase modulations or many stages) causes false generation of lower frequencies in the digital simulation environment. We simulate 10 independent sets of 256 symbols, as 256 symbols is a suitable block length for real-time continuous operation when considering the computational complexity in solving the transformation matrix~\cite{saxena2023performance}.

Throughout this paper we define the amount of dispersion per stage using the group delay dispersion $\beta_2 L$ (units of s$^2$), which is related to the typical dispersion parameter $D_2$ (typical units of ps/nm/km) by 
\begin{equation}
    \beta_2 L = -\frac{D_2 \lambda^2}{2 \pi c} L
\end{equation}
for device length $L$. This means that our results are independent of wavelength. We further normalise this to the symbol period $T_s$, resulting in units of $T_s^2$ for the dispersion quantities used throughout this paper $\beta_2 L$. This generalises the results to arbitrary baud rates/sampling rates. For example, for a device with 19.6~ps/nm dispersion, this results in a group delay dispersion of $\beta_2 L = 25$~ps$^2$ at 1550~nm. This is $T_s^2$ at 200~GBd symbol rate, since a 200~GBd signal has a syumbol duration of $T_s = 5$~ps.

The figure of merit used in this paper is signal-to-distortion ratio (SDR). This is simply the inverse of (\ref{DSR_definition}) 
\begin{equation}\label{SDR_definition}
    \textnormal{SDR} = \frac{1}{\textnormal{DSR}} = \frac{1}{\frac{1}{E_\textnormal{target}}\sum_{m=0}^M |\Psi_\textnormal{target}(mT) - \Psi_N(mT)|^2}
\end{equation}
and is fully equivalent to the commonly used signal-to-noise ratio (SNR) for communications channel characterisation. However we use the term `distortion' instead of `noise' to emphasise the fact that the deviation from the target waveform introduced by the limited fidelity of the transform is not random additive noise but rather a distortive property of the nonlinear transform. The distribution of these distortions is therefore unlikely to be white (i.e. equally likely across frequencies) or Gaussian. 

Fig.~\ref{fig:dispersion} shows how the signal-to-distortion ratio (SDR) varies with the dispersion per stage for $N = 1,\dots6$. In general, the more number of stages available, the higher SDR can be achieved. At any given number of stages $N$, there is a maximum SDR even if $\beta_2 L$ can be increased indefinitely.  This is because the function of the dispersive element is to provide sufficient temporal mode mixing to approximate the required non-zero matrix operation, $H$~\cite{borevich1981subgroups}, which is achieved at a finite amount of dispersion.  Overall, the distortion level can be very low (e.g. -33~dB) if a few (e.g. 5) stages with a small amount of dispersion per stage (e.g. $\beta_2 L =0.5~T_s^2$) can be implemented. For instance, at 200~GBd, $\beta_2 L =0.5~T_s^2$ is a dispersion of 9.8~ps/nm, which can easily be achieved on chip in standard silicon photonics~\cite{stern2023silicon}. In the extreme case of $N=1$, the SDR is around 3~dB does not change with dispersion since in this case the dispersive efficiency of a single element is defined purely by the target waveform~\cite{wyrowski1991upper,verhoeven2016upper}. The inset constellations highlight that when the amount of dispersion for a given $N$ is sub-optimal (Fig.~\ref{fig:dispersion}(a)), the main source of distortion is amplitude error, as is intuitive. On the contrary, when the SDR is limited by the number of stages (Fig.~\ref{fig:dispersion}(b)), the distortion appears equally distributed between phase and amplitude. Also inset are the power spectral densities of the RF waveforms for the phase modulators for every stage for two example cases: in this case of unrestricted power, the converged phase instructions show much stronger low frequency content and a strong roll off with increasing frequency.

\begin{figure}[t]
   \centering
    \input{figs/fig2}
    \caption{SDR v. phase modulator bandwidth ($B_{\textnormal{PM}}$), for $\beta_2 L =T_s^2$, and varying number of stages, $N$. The phase modulator bandwidth is plotted in units of symbol rate, $f_s$ and shaded areas indicate minimum and maximum values. Example constellations and power spectral densities of the converged phase modulations inset in (a) and (b).}
    \label{fig:pm_bw}

\end{figure}

Fig.~\ref{fig:pm_bw} shows that increasing the phase modulator bandwidth allows for more accurate modulation, even for the case of $N=1$ which saturates at the dispersive limit for a single phase only element~\cite{wyrowski1991upper}. Excess phase modulator bandwidth therefore allows access to high (e.g. $>30$ dB) SDRs even with lower (e.g. $N=3$) modulator counts. However, the important aspect of Fig.~\ref{fig:pm_bw} is that the nonlinear nature of the transform allows for the synthesis of waveforms far beyond the bandwidth of the constituent phase modulators. For example, 25~dB SDR is achievable with 6 stages of only $0.25~f_s$ modulator bandwidth. Therefore, a 400~GBd waveform could be generated with only 100~GHz modulator bandwidth, or a 200~GBd waveform with 50~GHz modulator bandwidth. Such a result is not surprising given well known results from temporal imaging theory~\cite{kolner1994space}, but this tradeoff between required modulator bandwidth and number of stages $N$ offers a design flexibility that is not available with linear IQ modulation, where the available bandwidth is strictly determined by the amplitude modulator bandwidth.

\begin{figure}[t]
   \centering
    \pgfmathsetmacro{\figuresizex}{1}
\pgfmathsetmacro{\figuresizey}{0.7}
\pgfmathsetmacro{\fillopacity}{0.2}
\pgfmathsetmacro{\figuresizeconst}{0.1}
\pgfmathsetmacro{\figuresizephasex}{0.19}
\pgfmathsetmacro{\figuresizephasey}{0.18}
\pgfmathsetmacro{\subfigsypos}{0.215}

\begin{tikzpicture}[trim axis left,trim axis right]

\begin{axis} [ylabel= SDR (dB), 
              xlabel={Block length (symbols)},
              xmin=4,xmax=1024,
              ymax=50,
              ymin=0,
              height=\figuresizey*\linewidth,
              grid=both,
              width=\figuresizex*\linewidth,
              cycle list/Dark2,
              ylabel near ticks,
              legend pos=north west,
              legend columns =3,
              clip mode=individual,
              log ticks with fixed point,
              legend style={font=\footnotesize},
              label style={font=\footnotesize},
              yticklabels={0,20,40,60,80,100,120,140,160},
              ytick={0,20,40,60,80,100,120,140,160},
              tick label style={font=\footnotesize},
              xmode=log,
              log basis x={2},
              xticklabels={4,16,64,256,1024,4096},
              xtick={4,16,64,256,1024,4096},]

\addplot+ [mark=*, thick] table [x=symbol_length,y=1, col sep=comma] {data/meanSNR_bit_sequence_length_v_N_stages__bits_10_QAM-16_100GBd_RRC_0.1_101taps_800GSa_dispersion_78_psnm_power_limit_None.csv} node[pos=0.7, sloped,anchor=south]{{\footnotesize $N=1$}};
\pgfplotsset{cycle list shift=-1}
\addplot+ [mark=none, thick, name path=A,draw opacity=\fillopacity] table [x=symbol_length,y=1, col sep=comma] {data/maxSNR_bit_sequence_length_v_N_stages__bits_10_QAM-16_100GBd_RRC_0.1_101taps_800GSa_dispersion_78_psnm_power_limit_None.csv};
\pgfplotsset{cycle list shift=-2}
\addplot+ [mark=none, thick, name path=B,draw opacity=\fillopacity] table [x=symbol_length,y=1, col sep=comma] {data/minSNR_bit_sequence_length_v_N_stages__bits_10_QAM-16_100GBd_RRC_0.1_101taps_800GSa_dispersion_78_psnm_power_limit_None.csv};
\pgfplotsset{cycle list shift=-3}
\addplot+[fill opacity=\fillopacity] fill between [of=A and B];

\addplot+ [mark=*, thick] table [x=symbol_length,y=2, col sep=comma] {data/meanSNR_bit_sequence_length_v_N_stages__bits_10_QAM-16_100GBd_RRC_0.1_101taps_800GSa_dispersion_78_psnm_power_limit_None.csv}node[pos=0.7, sloped,anchor=south]{{\footnotesize $N=2$}};
\pgfplotsset{cycle list shift=-4}
\addplot+ [mark=none, thick, name path=A,draw opacity=\fillopacity] table [x=symbol_length,y=2, col sep=comma] {data/maxSNR_bit_sequence_length_v_N_stages__bits_10_QAM-16_100GBd_RRC_0.1_101taps_800GSa_dispersion_78_psnm_power_limit_None.csv};
\pgfplotsset{cycle list shift=-5}
\addplot+ [mark=none, thick, name path=B,draw opacity=\fillopacity] table [x=symbol_length,y=2, col sep=comma] {data/minSNR_bit_sequence_length_v_N_stages__bits_10_QAM-16_100GBd_RRC_0.1_101taps_800GSa_dispersion_78_psnm_power_limit_None.csv};
\pgfplotsset{cycle list shift=-6}
\addplot+[fill opacity=\fillopacity] fill between [of=A and B];

\addplot+ [mark=*, thick] table [x=symbol_length,y=3, col sep=comma] {data/meanSNR_bit_sequence_length_v_N_stages__bits_10_QAM-16_100GBd_RRC_0.1_101taps_800GSa_dispersion_78_psnm_power_limit_None.csv}node[pos=0.7, sloped,anchor=south]{{\footnotesize $N=3$}};
\pgfplotsset{cycle list shift=-7}
\addplot+ [mark=none, thick, name path=A,draw opacity=\fillopacity] table [x=symbol_length,y=3, col sep=comma] {data/maxSNR_bit_sequence_length_v_N_stages__bits_10_QAM-16_100GBd_RRC_0.1_101taps_800GSa_dispersion_78_psnm_power_limit_None.csv};
\pgfplotsset{cycle list shift=-8}
\addplot+ [mark=none, thick, name path=B,draw opacity=\fillopacity] table [x=symbol_length,y=3, col sep=comma] {data/minSNR_bit_sequence_length_v_N_stages__bits_10_QAM-16_100GBd_RRC_0.1_101taps_800GSa_dispersion_78_psnm_power_limit_None.csv};
\pgfplotsset{cycle list shift=-9}
\addplot+[fill opacity=\fillopacity] fill between [of=A and B];

\addplot+ [mark=*, thick] table [x=symbol_length,y=4, col sep=comma] {data/meanSNR_bit_sequence_length_v_N_stages__bits_10_QAM-16_100GBd_RRC_0.1_101taps_800GSa_dispersion_78_psnm_power_limit_None.csv}node[pos=0.75, sloped,anchor=south]{{\footnotesize $N=4$}};
\pgfplotsset{cycle list shift=-10}
\addplot+ [mark=none, thick, name path=A,draw opacity=\fillopacity] table [x=symbol_length,y=4, col sep=comma] {data/maxSNR_bit_sequence_length_v_N_stages__bits_10_QAM-16_100GBd_RRC_0.1_101taps_800GSa_dispersion_78_psnm_power_limit_None.csv};
\pgfplotsset{cycle list shift=-11}
\addplot+ [mark=none, thick, name path=B,draw opacity=\fillopacity] table [x=symbol_length,y=4, col sep=comma] {data/minSNR_bit_sequence_length_v_N_stages__bits_10_QAM-16_100GBd_RRC_0.1_101taps_800GSa_dispersion_78_psnm_power_limit_None.csv};
\pgfplotsset{cycle list shift=-12}
\addplot+[fill opacity=\fillopacity] fill between [of=A and B];

\addplot+ [mark=*, thick] table [x=symbol_length,y=5, col sep=comma] {data/meanSNR_bit_sequence_length_v_N_stages__bits_10_QAM-16_100GBd_RRC_0.1_101taps_800GSa_dispersion_78_psnm_power_limit_None.csv}node[pos=0.8, sloped,anchor=south]{{\footnotesize $N=5$}};
\pgfplotsset{cycle list shift=-13}
\addplot+ [mark=none, thick, name path=A,draw opacity=\fillopacity] table [x=symbol_length,y=5, col sep=comma] {data/maxSNR_bit_sequence_length_v_N_stages__bits_10_QAM-16_100GBd_RRC_0.1_101taps_800GSa_dispersion_78_psnm_power_limit_None.csv};
\pgfplotsset{cycle list shift=-14}
\addplot+ [mark=none, thick, name path=B,draw opacity=\fillopacity] table [x=symbol_length,y=5, col sep=comma] {data/minSNR_bit_sequence_length_v_N_stages__bits_10_QAM-16_100GBd_RRC_0.1_101taps_800GSa_dispersion_78_psnm_power_limit_None.csv};
\pgfplotsset{cycle list shift=-15}
\addplot+[fill opacity=\fillopacity] fill between [of=A and B];

\addplot+ [mark=*, thick] table [x=symbol_length,y=6, col sep=comma] {data/meanSNR_bit_sequence_length_v_N_stages__bits_10_QAM-16_100GBd_RRC_0.1_101taps_800GSa_dispersion_78_psnm_power_limit_None.csv};
\pgfplotsset{cycle list shift=-16}
\addplot+ [mark=none, thick, name path=A,draw opacity=\fillopacity] table [x=symbol_length,y=6, col sep=comma] {data/maxSNR_bit_sequence_length_v_N_stages__bits_10_QAM-16_100GBd_RRC_0.1_101taps_800GSa_dispersion_78_psnm_power_limit_None.csv}node[pos=0.4, sloped,anchor=south]{{\footnotesize $N=6$}};
\pgfplotsset{cycle list shift=-17}
\addplot+ [mark=none, thick, name path=B,draw opacity=\fillopacity] table [x=symbol_length,y=6, col sep=comma] {data/minSNR_bit_sequence_length_v_N_stages__bits_10_QAM-16_100GBd_RRC_0.1_101taps_800GSa_dispersion_78_psnm_power_limit_None.csv};
\pgfplotsset{cycle list shift=-18}
\addplot+[fill opacity=\fillopacity] fill between [of=A and B];

\end{axis}
\end{tikzpicture}
    \caption{SDR v. block length in symbols, for $\beta_2 L =T_s^2$, $B_{\textnormal{PM}}=$ 0.55~$f_s$, and varying number of stages, $N$. Shaded areas indicate minimum and maximum values. }
    \label{fig:length}

\end{figure}
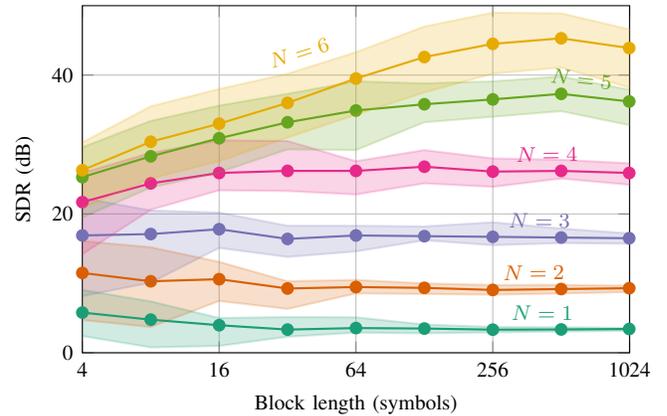

In Fig.~\ref{fig:length} we show how the number of symbols, i.e the block length limits the achievable SDR. We see that a low number of symbols ($<64$ in this case) leads to a lower SDR and greater uncertainty in the converged SDR. This is expected, since the modulation scheme relies on redistribution of energy to achieve amplitude modulations, and so sufficient time is needed to achieve this energy redistribution. 

A longer block length however, will clearly affect the computational complexity of the the optimisation procedure. Note that while decreasing the block length reduces the computational complexity of a single block, it is more relevant to consider the complexity per unit time. For example, the generalised L-BFGS algorithm used here has complexity $O(kW)$, for $k$ past updates used to calculate the Hessian and $W$ variables. In our case the number of variables is equal to the number of stages $N$ time the block length $M$,
\begin{equation}
    W = NM
\end{equation}
leading to the complexity of single block being $O(kNM)$. Since the complexity with respect to $M$ scales linearly, complexity per unit time is simply $O(kN)$. 

On the other hand, calculation of both the DSR (\ref{DSR_unconstrained}) and gradient (\ref{gradient_unconstrained}) functions, which is required at every iteration of the optimisation, requires the calculation of $3N$ FFT/IFFT pairs for $N$ stages: $N$ FFT/IFFT pairs for both the forward and backward propagating waves and a third set of $N$ FFT/IFFT pairs to impose the phase instructions filter constraint. Since the block length $M$ equals the FFT length, the calculation of each FFT/IFFT pair including a multiplication in the frequency domain requires $M\log_2(M) + M$ complex multiplications~\cite{spinnler2009complexity}, which for context is the same complexity as a static frequency domain equalizer. This results in a total of $3N(M \log_2(M)+M)$ complex multiplications per iteration of the optimisation algorithm, or $3N (\log_2(M) +1)$ per unit time. A reduction in block length $M$ can therefore reduce the complexity per unit time of the optimisation, although this will be small for large $M$ as it scales with $\log_2(M)$. In any case, the computational complexity of the optimisation procedure is a current drawback of this scheme compared to IQ modualtion, as was discussed previously in~\cite{saxena2023performance}.

Another issue is that continuous operation requires concatenating consecutive blocks of symbols. This results in degrading SDR at the block boundaries due to cyclic contamination of the dispersion impulse response. This issue can be eliminated by padding each block with symbols from the neighbouring block during optimisation, which are then discarded before modulation~\cite{saxena2023performance}.

\subsection{Constrained RF drive power}
\begin{figure*}[tb]
   \centering
    \input{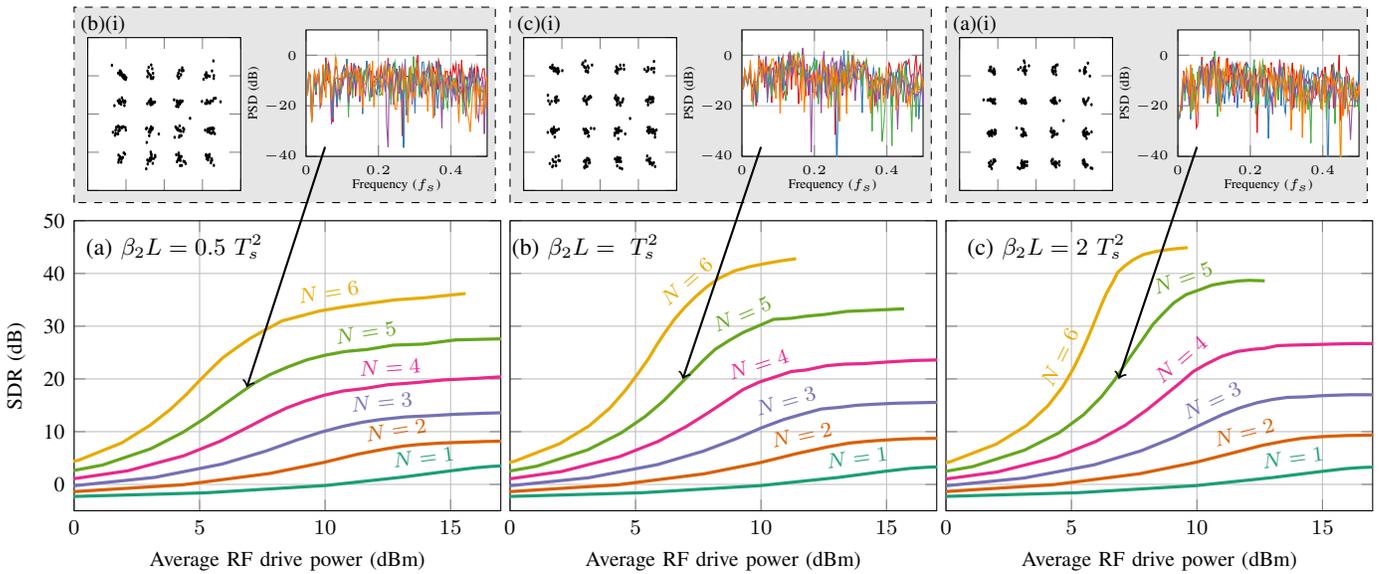}
    \caption{SDR v. average power for a $B_{\textnormal{PM}}=$ 0.55~$f_s$, and varying number of stages, $N$ for (a) $\beta_2 L =0.5~T_s^2$, (b) $\beta_2 L =T_s^2$ and (c) $\beta_2 L =2~T_s^2$ The dispersion is normalised to the symbol period squared ($T_s^2$). We assume $V_\pi = 3$~V and 50~Ohm impedance for the average power calculation. }
    \label{fig:power_constraint}

\end{figure*}
The results for multi-objective optimization to simultaneously optimise drive power are shown in Fig.~\ref{fig:power_constraint}. We simulate 3 different dispersion values, (a) $\beta_2 L =0.5~T_s^2$, (b) $\beta_2 L =T_s^2$ and (c) $\beta_2 L =2~T_s^2$ and minimize the objective function (\ref{multi_objective}) for a range of scalarization parameters in the range $a = [10^{-5}, 10^{0.5}]$. We then calculate the average power of the drive instructions per stage under the assumption that the phase modulator has $V_\pi = 3$~V and 50~Ohm impedance.  

In general, Fig.~\ref{fig:power_constraint} shows that increasing the available power increases achievable SDR at a given number of stages until the SDR gradually saturates at the maximum SDR given by the results in Fig.~\ref{fig:dispersion} and Fig.~\ref{fig:pm_bw}. Roughly around 10~dBm average power is needed substantially reduce the penalty from restricting the driver power, with minor gains of $<5$~dB SDR seen up to around 15 dBm where the SDR matches that of the unconstrained power optimisation results. The typical output power of SiGe drivers in coherent optical products is around 10~dBm~\cite{torfs2023high} which highlights the suitability of implementing this scheme using existing driver technology. The required  average power could be substantially reduced by advances in modulator technology reducing $V_\pi$ such as thin film lithium niobate~\cite{wang2018integrated}. Unlike conventional IQ modulation however, this approach introduces more system design flexibility by allowing the possibility of reducing the required driver power by adding more modulation stages. This could enable driving the modulation directly from the CMOS DAC/ASIC, eliminating the integration complexity and cost of using SiGe driver technology between the CMOS ASIC and photonic integrated circuit.

Note also that any penalties associated with reducing the drive power may also be compensated by increasing the dispersion (see increase in SDR from Fig.~\ref{fig:power_constraint}(a) to Fig.~\ref{fig:power_constraint}(c)) until reaching the maximum SDR values for the unconstrained power case in Fig.~\ref{fig:dispersion}, the requirements for which will reduce quadratically with an increase in baudrate. The opposite is also true: in a situation where on-chip dispersion is limited increasing the drive power may improve the SDR. All of these trade-offs emphasise the design flexibility of the spectro-temporal unitary transform scheme.

One important difference between Fig.~\ref{fig:power_constraint} and the optimisation without power constraints in the previous figures (Figs.~\ref{fig:dispersion},~\ref{fig:pm_bw} and~\ref{fig:length}) is that the phase modulations power spectral densities, as shown in the insets (Fig.~\ref{fig:power_constraint}(a)(i), Fig.~\ref{fig:power_constraint}(b)(i), Fig.~\ref{fig:power_constraint}(c)(i)), are much flatter with respect to frequency. This shows that when the RF power is constrained to realistic levels, the spectral densities of the drive signals is similar to that of conventional IQ modulation, and so should be compatible with broadband integrated drivers. 

Finally, it is important to emphasise that the distortions discussed in this section are not additive white Gaussian noise (AWGN). As with other nonlinear channels~\cite{kramer2015upper,essiambre2010capacity,yousefi2014information}, the resulting distortions may affect the forward error correction (FEC) and digital signal processing functions in a substantially different way to the auxiliary AWGN channel for which they are typically designed. The relationship between capacity (i.e. mutual information) and SDR for this modulation scheme is not the focus of this paper but would be important to establish to understand the true capacity limits. 

\subsection{Required optical power}

Although the spectro-temporal unitary transformation is theoretically lossless, the phase modulators and dispersive elements will in practice have insertion loss. Therefore in Fig.~\ref{fig:laser_power} we calculate the transmitter SINAD resulting from the combination of transform limited SDR and the optical SNR (OSNR) as limited by shot noise for a given input laser power. We use an example of 200~GBd symbol rate (i.e. 200~GHz transmitter optical bandwidth), with $B_{\textnormal{PM}} =110$~GHz, $19.6~$~ps/nm dispersion ($\beta_2L = T_s^2$) and assume each stage has a realistic 1-dB insertion loss, resulting from approximately 0.5~dB each from the modulator and dispersive element\cite{zhu2021integrated,stern2023silicon}. The transform limited SDR values are taken from Fig.~\ref{fig:power_constraint}(b), for an RF drive power of 10~dBm. We also calculate modulation loss for the conventional IQ case from the actual Nyquist-shaped 16-QAM electrical waveform driving the sinusoidal MZM transfer function with a modulation depth of $V_{\textnormal{peak}}/V_\pi = 0.25$. The modulation depth is chosen for linear operation~\cite{chen201916384}, resulting in an average modulation loss of 21~dB. 

The SINAD for the conventional IQ modulation case is indicated by the dashed line in Fig.~\ref{fig:power_constraint}(b) and is simply the OSNR after modulation. At all laser powers, Fig.~\ref{fig:laser_power}(b) shows that the the low loss nature of the transformation allows it to achieve substantially higher SINADs than the conventional IQ modulation case ($\approx 15$~dB higher), provided enough stages are implemented. Fewer stages results in lower modulation loss, although this effect is small given the low loss of each stage (1~dB) and would only become relevant if a large number of stages was implemented. An important point to emphasize is that Fig.~\ref{fig:laser_power} is plotted for 200~GBd: as the baudrate is increased, shot noise will increase proportionally, leading to an corresponding decrease in OSNR. For example, increasing to 500~GBd (approximate symbol rate for a hypothetical 3200ZR/ZR+ transceiver) would decrease the SINAD of IQ modulation by 4~dB and therefore below the typical required 35~dB OSNR for optical transceiver, considering the power range of current integrable tunable laser assemblies (indicated by red shaded area in Fig.~\ref{fig:laser_power}). The low loss unitary transform technique therefore becomes increasingly attractive at higher symbol rates, assuming the transform SDR can be maintained. 

\begin{figure}[t]
   \centering
    \begin{tikzpicture}[trim axis left,trim axis right]

\begin{axis} [ylabel= SINAD (dB), 
              xlabel=Laser power  (dBm),
              xmin=-40,xmax=20,
              ymax=50,
              ymin=-10,
              height=0.7*\linewidth,
              grid=both,
              width=\linewidth,
              cycle list/Dark2,
              ylabel near ticks,
              legend columns =3,
              clip mode=individual,
              log ticks with fixed point,
              legend style={font=\footnotesize},
              legend pos=north west,
              label style={font=\footnotesize},
              yticklabels={-10,0,10,20,30,40,50,60,70},
              ytick={-10,0,10,20,30,40,50,60,70},
              tick label style={font=\footnotesize},]

\addplot+[draw=none,fill=red!15,forget plot] coordinates {(13,-10) (13,50) (17,50) (17,-10)};

\addplot [mark=none, very thick,black,dashed] table [x=laser_powers,y=IQ, col sep=comma] {data/laser_power_1dB_loss_0.25_Vp_200GBd.csv} node[pos=0.32, sloped,anchor=north]{{}};
\addlegendentry{IQ mod.}

\addplot+ [mark=none, very thick] table [x=laser_powers,y=2, col sep=comma] {data/laser_power_1dB_loss_0.25_Vp_200GBd.csv} node[pos=0.9, sloped,anchor=south]{{\footnotesize $N=2$}};

\pgfplotsset{cycle list shift=-2}

\addplot+ [mark=none, very thick] table [x=laser_powers,y=1, col sep=comma] {data/laser_power_1dB_loss_0.25_Vp_200GBd.csv} node[pos=0.6, sloped,anchor=south,yshift=-2]{{\footnotesize $N=1$}};
\pgfplotsset{cycle list shift=-1}
\addplot+ [mark=none, very thick] table [x=laser_powers,y=3, col sep=comma] {data/laser_power_1dB_loss_0.25_Vp_200GBd.csv} node[pos=0.7, sloped,anchor=south]{{\footnotesize $N=3$}};

\addplot+ [mark=none, very thick] table [x=laser_powers,y=4, col sep=comma] {data/laser_power_1dB_loss_0.25_Vp_200GBd.csv} node[pos=0.83, sloped,anchor=south]{{\footnotesize $N=4$}};

\addplot+ [mark=none, very thick] table [x=laser_powers,y=5, col sep=comma] {data/laser_power_1dB_loss_0.25_Vp_200GBd.csv} node[pos=0.77, sloped,anchor=south]{{\footnotesize $N=5$}};

\addplot+ [mark=none, very thick] table [x=laser_powers,y=6, col sep=comma] {data/laser_power_1dB_loss_0.25_Vp_200GBd.csv} node[pos=0.88, sloped,anchor=south]{{\footnotesize $N=6$}};

\draw[thick,<->] (axis cs:13,-3) -- (axis cs:17,-3) node[pos=0.2, sloped,anchor=north]{{\footnotesize ITLA power}};

\end{axis}

\end{tikzpicture}
    \caption{SINAD v laser power for 200 GBd, assuming a modulation loss per stage of 1~dB, $B_{\textnormal{PM}} =110$~GHz, $19.6~$~ps/nm dispersion. For the IQ mod. case, we assume a modulation depth of $V_{\textnormal{peak}}/V_\pi = 0.25$. The red shaded area indicates typical power range (13 to 17 dBm) of an integrable tunable laser assembly (ITLA).}
    \label{fig:laser_power}

\end{figure}

\section{Practical limitations (System Tolerance to Fabrication Errors)}

The simulations in the previous section assumed that phase instructions can be modulated onto the optical signal with perfect accuracy. Since the transform scheme requires multiple stages of nonlinear modulation, errors in amplitude or phase do not propagate through the modulator in linear fashion like a conventional IQ modulation. In addition, errors in the fabricated dispersive element may further contribute to SDR degradation since the phase instructions are calculated assuming a specific linear group delay dispersion value. Therefore, it is important to understand how these potential errors degrade the SDR and the resulting require calibration of any fabricated device. 

The calculated phase instructions $\{\phi_1,\phi_2,\dots \phi_N\}$ as applied to the optical waveform may actually be
\begin{equation}
    \phi_{n,\textnormal{err}} = \alpha\phi_n(t+ \psi) 
\end{equation}

 $\alpha$ is a fixed amplitude error that originate for example from incorrect driver output amplitude or unexpected loss in the signal path. $\psi$ is a fixed phase error that could result for example from unaccounted for delays in the electrical signal driving path, timing errors, or delays in the optical waveguide between stages. These fixed errors are in theory correctable through calibration. 

\begin{figure}[t]
   \centering
    \begin{tikzpicture}
    \begin{semilogxaxis}[
                ylabel= SDR (dB), 
              xlabel={Phase error ($T_s$)},
              xmin=10^-5,xmax=1,
              ymax=45,
              ymin=-5,
              height=0.7*\linewidth,
              grid=both,
              width=\linewidth,
              cycle list/Dark2,
              ylabel near ticks,
              legend pos=north west,
              legend columns =3,
              clip mode=individual,
              legend style={font=\footnotesize},
              label style={font=\footnotesize},
              yticklabels={0,10,20,30,40,50,60,70},
              ytick={0,10,20,30,40,50,60,70},
              tick label style={font=\footnotesize}]
              
        \addplot+ [mark=none, very thick] table [x=error,y=1, col sep=comma] {data/phase_error_analysis__bits_10_QAM-16_100GBd_RRC_0.1_101taps_800GSa_dispersion_78_psnm_power_limit_None.csv} node[pos=0.1, sloped,anchor=south]{{\footnotesize $N=1$}};

        \addplot+ [mark=none, very thick] table [x=error,y=2, col sep=comma] {data/phase_error_analysis__bits_10_QAM-16_100GBd_RRC_0.1_101taps_800GSa_dispersion_78_psnm_power_limit_None.csv} node[pos=0.1, sloped,anchor=south]{{\footnotesize $N=2$}};

        \addplot+ [mark=none, very thick] table [x=error,y=3, col sep=comma] {data/phase_error_analysis__bits_10_QAM-16_100GBd_RRC_0.1_101taps_800GSa_dispersion_78_psnm_power_limit_None.csv} node[pos=0.1, sloped,anchor=south]{{\footnotesize $N=3$}};

        \addplot+ [mark=none, very thick] table [x=error,y=4, col sep=comma] {data/phase_error_analysis__bits_10_QAM-16_100GBd_RRC_0.1_101taps_800GSa_dispersion_78_psnm_power_limit_None.csv} node[pos=0.05, sloped,anchor=south]{{\footnotesize $N=4$}};

        \addplot+ [mark=none, very thick] table [x=error,y=5, col sep=comma] {data/phase_error_analysis__bits_10_QAM-16_100GBd_RRC_0.1_101taps_800GSa_dispersion_78_psnm_power_limit_None.csv} node[pos=0.05, sloped,anchor=south]{{\footnotesize $N=5$}};

        \addplot+ [mark=none, very thick] table [x=error,y=6, col sep=comma] {data/phase_error_analysis__bits_10_QAM-16_100GBd_RRC_0.1_101taps_800GSa_dispersion_78_psnm_power_limit_None.csv} node[pos=0.05, sloped,anchor=south]{{\footnotesize $N=6$}};
        
    \end{semilogxaxis}
\end{tikzpicture}
    \caption{Inter-stage phase error v. achievable SDR for $\beta_2 L =~T_s^2$, $B_{\textnormal{PM}}=$ 0.55~$f_s$, and varying number of stages, $N$. It is assumed that each stage experiences the same error.}
    \label{fig:phase_error}

\end{figure}
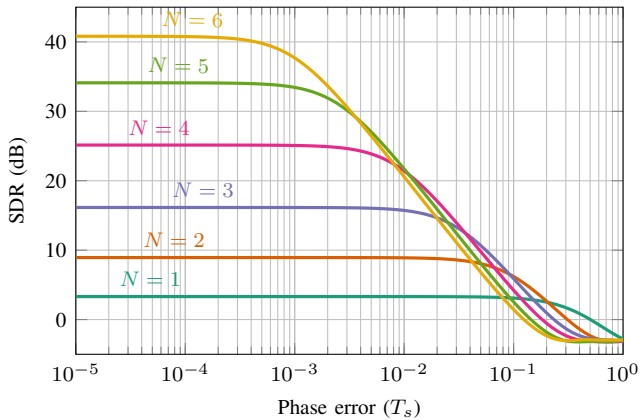

Fig.~\ref{fig:phase_error} shows how the phase error $\psi$ reduces the overall SDR on an example RRC-shaped 16-QAM waveform with $\beta_2 L =T_s^2$, $B_{\textnormal{PM}}=$ 0.55~$f_s$, and varying number of stages, $N$. For these results, we assumed that every stage experiences the same phase delay, e.g. due to systematic fabrication error. In reality, it is unlikely that each stage would experience the same error but it is instructive for understanding the scale of the penalities associated with a fixed phase error. Furthermore, this fixed error on every stage is equivalent to a random error with standard deviation equal to the fixed error. 

Notwithstanding the impact of the number of stages on SDR, a higher number of stages generally results in a larger SDR degradation as is clear in high phase error scenarios (e.g. $10^{-1}T_s$). This is simply because the phase accuracy must be maintained across more stages. For small phase errors however, this effect is small and the SDR is mostly defined simply by the phase error, irrespective of the number of stages. For example, to achieve an SDR of 25~dB a phase error of around $5\times10^{-3}~T_s$ is required. At 200~GBd ($T_s = 5$~ps), this represents a time delay of 25~fs which corresponds to a length of 7.5~um in free space, 3.41~um in lithium niobate ($n\approx2.2$) or 2.14~um in silicon ($n\approx3.5$). Achieving this accuracy of path length matching on a photonic integrated circuit is trivial, which typically have minimum feature sizes on the order of $\approx 100$~nm, and so no calibration would be required on a well designed and fabricated die. With these assumptions, a fabrication accuracy of around $\approx 100$~nm would not cause significant SDR degradation until the symbol rate reaches $>10$~TBd. In addition, the delay deviation due to fabrication variation in waveguide width is generally lower than 10 fs based on typical silicon foundry cross-wafer statistics\cite{siew2021review}. Besides optical path length mismatch, electrical timing skew may also cause the errors plotted in Fig.~\ref{fig:phase_error}, either in the RF chain to the modulators or from digital timing effects. As with any optical path length mismatch, a well designed chip or calibration routine should keep these errors negligible.

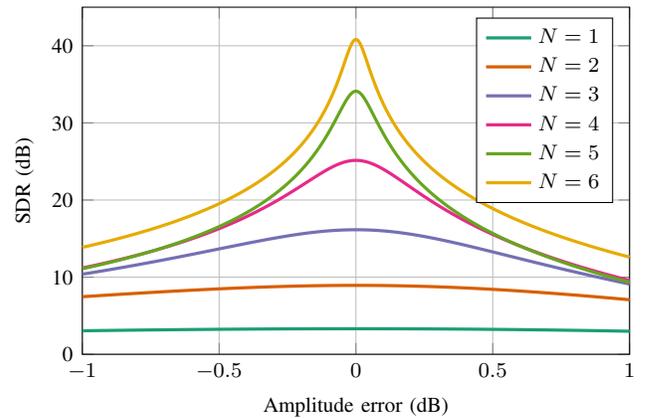
\begin{figure}[t]
   \centering
    \begin{tikzpicture}
    \begin{axis}[
                ylabel= SDR (dB), 
              xlabel={Amplitude error (dB)},
              xmin=-1,xmax=1,
              ymax=45,
              ymin=0,
              height=0.7*\linewidth,
              grid=both,
              width=\linewidth,
              cycle list/Dark2,
              ylabel near ticks,
              legend pos=north east,
              legend columns =1,
              clip mode=individual,
              legend style={font=\footnotesize},
              label style={font=\footnotesize},
              yticklabels={0,10,20,30,40,50,60,70},
              ytick={0,10,20,30,40,50,60,70},
              tick label style={font=\footnotesize}]
              
        \addplot+ [mark=none, very thick] table [x=error,y=1, col sep=comma] {data/amp_error_analysis__bits_10_QAM-16_100GBd_RRC_0.1_101taps_800GSa_dispersion_78_psnm_power_limit_None.csv};
        \addlegendentry{$N=1$}

        \addplot+ [mark=none, very thick] table [x=error,y=2, col sep=comma] {data/amp_error_analysis__bits_10_QAM-16_100GBd_RRC_0.1_101taps_800GSa_dispersion_78_psnm_power_limit_None.csv};
        \addlegendentry{$N=2$}

        \addplot+ [mark=none, very thick] table [x=error,y=3, col sep=comma] {data/amp_error_analysis__bits_10_QAM-16_100GBd_RRC_0.1_101taps_800GSa_dispersion_78_psnm_power_limit_None.csv};
        \addlegendentry{$N=3$}

        \addplot+ [mark=none, very thick] table [x=error,y=4, col sep=comma] {data/amp_error_analysis__bits_10_QAM-16_100GBd_RRC_0.1_101taps_800GSa_dispersion_78_psnm_power_limit_None.csv};
        \addlegendentry{$N=4$}

        \addplot+ [mark=none, very thick] table [x=error,y=5, col sep=comma] {data/amp_error_analysis__bits_10_QAM-16_100GBd_RRC_0.1_101taps_800GSa_dispersion_78_psnm_power_limit_None.csv};
        \addlegendentry{$N=5$}

        \addplot+ [mark=none, very thick] table [x=error,y=6, col sep=comma] {data/amp_error_analysis__bits_10_QAM-16_100GBd_RRC_0.1_101taps_800GSa_dispersion_78_psnm_power_limit_None.csv};
        \addlegendentry{$N=6$}
    \end{axis}
\end{tikzpicture}
    \caption{Phase instruction drive power error v. SDR for $\beta_2 L = T_s^2$, $B_{\textnormal{PM}}=$ 0.55~$f_s$, and varying number of stages, $N$. It is assumed that each stage experiences the same error.}
    \label{fig:amp_error}

\end{figure}

The SDR degradation caused by fixed amplitude errors is show in Fig.~\ref{fig:amp_error}. As with the phase error in Fig.~\ref{fig:phase_error}, we apply the same amplitude error to every stage simply by scaling the phase instructions by the factor $\alpha$. Unlike Fig.~\ref{fig:phase_error}, the amplitude error SDR degradation does not increase with an increasing number of stages $N$. For example, $N=6$ maintains the highest SDR for all amplitude error values. However, it is seen that the spectro-temporal unitary transform does experience strong sensitivity to the amplitude accuracy of the phase transformations: for example, achieving an SDR $>30$~dB requires an power accuracy of around 0.05~dB, or around a 0.5\% accuracy in amplitude. Such accuracy in generally achievable in CMOS DACs but will require careful calibration routines. It should also be noted that the amplitude error experiences asymmetry in SDR degradation due to the nonlinear nature of phase modulation: i.e. the positive amplitude errors ($>0$~dB) in Fig.~\ref{fig:amp_error} have a lower SDR that their corresponding negative errors ($<0$~dB). This is most evident in large error values (e.g. $\pm 1$~dB) in Fig.~\ref{fig:amp_error}.

\begin{figure}[t]
   \centering
    \begin{tikzpicture}
    \begin{axis}[
                ylabel= SDR (dB), 
              xlabel={Dispersion error ($T_s^2$)},
              xmin=-0.05,xmax=0.05,
              ymax=45,
              ymin=0,
              height=0.7*\linewidth,
              grid=both,
              width=\linewidth,
              cycle list/Dark2,
              ylabel near ticks,
              legend pos=north east,
              legend columns =1,
              clip mode=individual,
              legend style={font=\footnotesize},
              label style={font=\footnotesize},
              yticklabels={0,10,20,30,40,50,60,70},
              ytick={0,10,20,30,40,50,60,70},
              scaled ticks=false,
              tick label style={font=\footnotesize,/pgf/number format/fixed}]
              
        \addplot+ [mark=none, very thick] table [x=error,y=1, col sep=comma] {data/disp_error_analysis__bits_10_QAM-16_100GBd_RRC_0.1_101taps_800GSa_dispersion_-78_psnm_power_limit_None.csv};
        \addlegendentry{$N=1$}

        \addplot+ [mark=none, very thick] table [x=error,y=2, col sep=comma] {data/disp_error_analysis__bits_10_QAM-16_100GBd_RRC_0.1_101taps_800GSa_dispersion_-78_psnm_power_limit_None.csv};
        \addlegendentry{$N=2$}

        \addplot+ [mark=none, very thick] table [x=error,y=3, col sep=comma] {data/disp_error_analysis__bits_10_QAM-16_100GBd_RRC_0.1_101taps_800GSa_dispersion_-78_psnm_power_limit_None.csv};
        \addlegendentry{$N=3$}

        \addplot+ [mark=none, very thick] table [x=error,y=4, col sep=comma] {data/disp_error_analysis__bits_10_QAM-16_100GBd_RRC_0.1_101taps_800GSa_dispersion_-78_psnm_power_limit_None.csv};
        \addlegendentry{$N=4$}

        \addplot+ [mark=none, very thick] table [x=error,y=5, col sep=comma] {data/disp_error_analysis__bits_10_QAM-16_100GBd_RRC_0.1_101taps_800GSa_dispersion_-78_psnm_power_limit_None.csv};
        \addlegendentry{$N=5$}

        \addplot+ [mark=none, very thick] table [x=error,y=6, col sep=comma] {data/disp_error_analysis__bits_10_QAM-16_100GBd_RRC_0.1_101taps_800GSa_dispersion_-78_psnm_power_limit_None.csv};
        \addlegendentry{$N=6$}
    \end{axis}
\end{tikzpicture}
    \caption{Linear dispersion error v. SDR for $\beta_2 L = T_s^2$, $B_{\textnormal{PM}}=$ 0.55~$f_s$, and varying number of stages, $N$. It is assumed that each stage experiences the same error.}
    \label{fig:disp_error}

\end{figure}
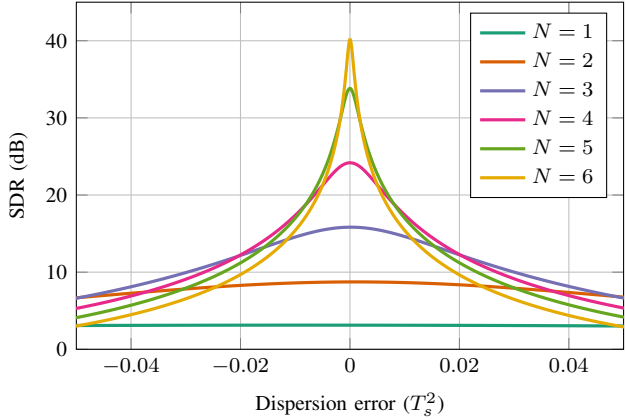

Achievable SDR as a function of linear dispersive error is shown in Fig.~\ref{fig:disp_error}. In general, the SDR is quite sensitive to dispersion error, requiring around $0.0017~T_s^2$ dispersion accuracy to achieve a SDR of $>30$~dB. At 200~GBd, 1550~nm for example, this represents a dispersion accuracy of around $\pm0.03$~ps/nm for a target dispersion value of $T_s^2, \beta_2 L = 19.6$~ps/nm. Although this requirement is quite stringent, the optimisation procedure can always be adjusted to account for the actual fabricated dispersion, as long as sufficiently good calibration routines can be established. This is true even if the dispersion has higher order elements, such as group delay ripples as a function of frequency. When the dispersive error is high, increasing the number of stages generally increases the SDR error, since the error is propagated across more stages. Significant higher order dispersion may however change the achievable SDR for a given configuration, but this analysis is outside of the scope of this paper in which we seek to establish general trends.  

\begin{figure}[t]
   \centering
    \pgfmathsetmacro{\figuresizex}{0.5}
\pgfmathsetmacro{\figuresizey}{0.29}

\begin{tikzpicture}[trim axis left,trim axis right]

\begin{axis} [ylabel= SINAD (dB), 
              xlabel={DAC resolution (bits)},
              xmin=1,xmax=10,
              ymax=60,
              ymin=-5,
              height=0.7\linewidth,
              grid=both,
              width=\linewidth,
              cycle list/Dark2,
              ylabel near ticks,
              legend pos=north west,
              legend columns =3,
              clip mode=individual,
              log ticks with fixed point,
              legend style={font=\footnotesize},
              label style={font=\footnotesize},
              yticklabels={0,10,20,30,40,50,60,70},
              ytick={0,10,20,30,40,50,60,70},
              xticklabels={1,2,3,4,5,6,7,8,9,10,11,12},
              xtick={1,2,3,4,5,6,7,8,9,10,11,12},
              tick label style={font=\footnotesize},]

\addplot+ [mark=none, very thick] table [x=bits,y=1, col sep=comma] {data/meanSNR_ENOB_calc__bits_10_QAM-16_100GBd_RRC_0.1_101taps_800GSa_BW_55_GHz_power_limit_None.csv} node[pos=0.95, sloped,anchor=north]{{\footnotesize $N=1$}};

\addplot+ [mark=none, very thick] table [x=bits,y=2, col sep=comma] {data/meanSNR_ENOB_calc__bits_10_QAM-16_100GBd_RRC_0.1_101taps_800GSa_BW_55_GHz_power_limit_None.csv} node[pos=0.95, sloped,anchor=south]{{\footnotesize $N=2$}};

\addplot+ [mark=none, very thick] table [x=bits,y=3, col sep=comma] {data/meanSNR_ENOB_calc__bits_10_QAM-16_100GBd_RRC_0.1_101taps_800GSa_BW_55_GHz_power_limit_None.csv} node[pos=0.96, sloped,anchor=south]{{\footnotesize $N=3$}};

\addplot+ [mark=none, very thick] table [x=bits,y=4, col sep=comma] {data/meanSNR_ENOB_calc__bits_10_QAM-16_100GBd_RRC_0.1_101taps_800GSa_BW_55_GHz_power_limit_None.csv} node[pos=0.98, sloped,anchor=south]{{\footnotesize $N=4$}};

\addplot+ [mark=none, very thick] table [x=bits,y=5, col sep=comma] {data/meanSNR_ENOB_calc__bits_10_QAM-16_100GBd_RRC_0.1_101taps_800GSa_BW_55_GHz_power_limit_None.csv} node[pos=0.98, sloped,anchor=south]{{\footnotesize $N=5$}};

\addplot+ [mark=none, very thick] table [x=bits,y=6, col sep=comma] {data/meanSNR_ENOB_calc__bits_10_QAM-16_100GBd_RRC_0.1_101taps_800GSa_BW_55_GHz_power_limit_None.csv} node[pos=0.98, sloped,anchor=south]{{\footnotesize $N=6$}};

\addplot+[black, dashed,very thick, domain=0:10]{6.02*x + 1.76} node[pos=0.3, sloped,anchor=south]{{\footnotesize SINAD = $6.02M + 1.76$}};

\end{axis}

\end{tikzpicture}
    \caption{DAC resolution v. SINAD, for $\beta_2 L =T_s^2$, $B_{\textnormal{PM}}=$ 0.55~$f_s$, and varying number of stages, $N$. The phase modulations are optimised with a power constraint such that $a=10^{-4}$,}
    \label{fig:DAC}

\end{figure}
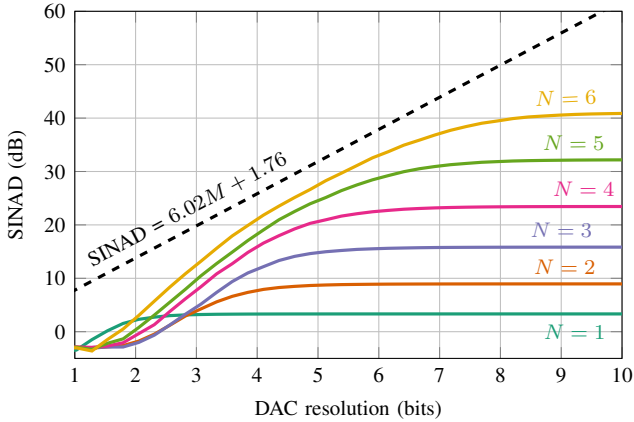

Finally, we assess the impact of DAC quantisation errors on the signal SDR. To do this we quantize the previously optimised phase instructions for the case $\beta_2 L = T_s^2$, $B_{\textnormal{PM}}=$ 0.55~$f_s$ with power constraint $a=10^{-4}$ into $2^{X}$ levels for $X$ DAC bits and plot the overall SINAD in Fig.~\ref{fig:DAC}. At high resolution, the SINAD simply saturates at the transform limited SDRs. However, at low resolution, the SINAD is substantially lower than the theoretical limit ($6.02M + 1.76$, for $M$ bits), which is a result of the nonlinear relationship between the successive phase modulation stages. For example, for $N=6$ stages a DAC resolution of 4 bits results in approximately 21~dB SINAD when the resolution limit of the DAC itself suggests a SINAD limit of approximately 26~dB. Nevertheless, Fig.~\ref{fig:DAC}, shows that a practical DAC resolution of $>6$ bits would offer a sufficiently low noise/distortion floor (e.g. $<-$30 dB). Note that lower bandwidth DACs in general can achieve higher resolution (effective number of bits, ENOB), and so it may be advantageous to use lower bandwidth, higher ENOB converters and more modualtion stages to achieve the highest possible overall SINAD.

\section{Conclusion}

We have modelled the achievable SDR for generating coherent modulation using spectro-temporal unitary transformations. This paper focussed on RRC shaped 16-QAM signals although the results are generally applicable to $N$-QAM signals. We have modelled how the number of stages, dispersion, phase modulator bandwidth and symbol block length effect the achievable SDR. Furthermore, we introduce a multi-objective optimisation procedure that also accounts for the limited modulator driver power available at the transmitter. Our results indicate that a high ($>30$~dB) SDR at $>200$~GBd is achievable with a low ($<6$) number of stages and reasonable parameters for driver power, modulator bandwidth and on-chip dispersion. Finally, we also investigated how errors in phase, amplitude, dispersion and limited DAC resolution reduce the SDR and establish quantitative guidelines on the required accuracy of these parameters to implement these transforms in a real chip-scale system.

\bibliographystyle{ieeetr}
\bibliography{sample}

\end{document}